\documentclass{iopart}
\newcommand{\bee}{\begin{equation}}
\newcommand{\eeq}{\end{equation}}

\def\la{\mathrel{\mathchoice {\vcenter{\offinterlineskip\halign{\hfil
$\displaystyle##$\hfil\cr<\cr\sim\cr}}}
{\vcenter{\offinterlineskip\halign{\hfil$\textstyle##$\hfil\cr<\cr\sim\cr}}}
{\vcenter{\offinterlineskip\halign{\hfil$\scriptstyle##$\hfil\cr<\cr\sim\cr}}}
{\vcenter{\offinterlineskip\halign{\hfil$\scriptscriptstyle##$\hfil\cr<\cr\sim\cr}}}}}

\newcommand{\bea}{\begin{eqnarray}}
\newcommand{\eea}{\end{eqnarray}}

\begin{document}
\jl{6}
\title{Combining general relativity and
quantum theory: points of conflict and contact}
\author{T. Padmanabhan}
\footnote[1]{E-mail address: {\tt nabhan@iucaa.ernet.in}}
\address{Inter-University Centre for Astronomy and Astrophysics, Post Bag 4, Ganeshkhind, Pune - 411 007.}
\date{}
\begin{abstract}
The issues related to bringing together the principles of general relativity and quantum theory are discussed. After briefly summarising the points of conflict between the two formalisms I focus on four specific themes in which some contact has been established in the past between GR and quantum field theory: (i) The role of planck length in the microstructure of spacetime (ii) The role of quantum effects in cosmology and origin of the universe (iii) The thermodynamics of spacetimes with horizons and especially the concept of entropy related to spacetime geometry (iv) The problem of the cosmological constant.
\end{abstract}


 
\section {Introduction}

The question of bringing together the principles of quantum theory
and gravity deserves to be called {\it the} problem  of theoretical physics
today. In this review I shall highlight the points of conflict {\it and} contact between these two theoretical structures focusing on four major themes
which run through all the work in quantum gravity over decades: (i) What is the role played
by the length scale 
$L_P\equiv (G\hbar/c^3)^{1/2}$ in determining the spacetime microstructure 
(see eg. \cite{wheel64} - \cite{tp97}) ? (ii) What have we learnt regarding the role of quantum gravity in quantum cosmology and in the origin of the universe
(see eg.\cite{dewitt67} - \cite{mbojo01} )? (iii) To what extent do we understand the thermodynamics of spacetimes with horizons 
(see eg. \cite{stein7274} - \cite{tp121124}) ? (iv) What is the role of quantum gravity vis-a-vis
the cosmological constant  (see eg. \cite{sakh68},\cite{ewit01}) ?
 Even among these themes, I will concentrate more on the latter two.
Before I discuss the concrete issues, it is probably worth comparing some general aspects of quantum theory and general relativity.

\section{ The miracle of quantum field theory}

The key feature of quantum field theory is that {\it it has no right to be as successful as it is!}.
In proceeding from classical mechanics [with finite number of
degrees of freedom] to quantum mechanics, one attributes
operator status to various dynamical variables and imposes the commutation relations among them.  Often, it is
convenient to provide a representation for the operators in terms of normal
differential operators so that the problem can be mapped to
solving a partial differential equation --- say, the time-dependent Schroedinger equation --- with specific boundary conditions. Such problems are mathematically well defined and tractable, allowing us to  construct
a well defined [though, in general, not unique] quantum theory for a classical
system with finite number of degrees of freedom.

The generalisation of such a procedure to a {\it field} with infinite number of
degrees of freedom is {\it not} straightforward. Given a classical field with some dynamical variables,  one can attempt to quantise
the system  by elevating the status of dynamical variables to operators
and imposing the commutation rules.
But finding a well defined and meaningful representation for this commutator algebra is a nontrivial task. Further, if one tries to extend the approach
of quantum mechanics [based on Schroedinger picture] to the field, one obtains
a {\it functional } differential equation instead of a partial differential
equation. The properties  --- let alone solutions! --- of this equation are
not well understood for any field with nontrivial interactions. Somewhat
simpler (and better) approach will be to use the Heisenberg picture and try to
solve for the operator valued distributions representing the various dynamical
variables. Even in this case, one does not have a systematic mathematical
machinery to solve these equations for an interacting field theory. The evolution equations for operators in QED [in 3+1 dimensions], for example,  cannot be solved exactly; however, it is possible to set up a perturbation expansion for
the relevant variables in powers of the coupling constant $(e^2/\hbar c)\approx 10^{-2}$. The lowest order of the perturbation series, in which all interactions
are switched off,  defines the {\it free field} theory which can be mapped to a model describing infinite number of noninteracting
harmonic oscillators.  The
perturbation expansion can be then used to obtain the ``corrections''
to this free
field theory. Several nontrivial conceptual issues crop up when such an attempt is made:

(a) To begin with, the decomposition of the field in terms of the harmonic
oscillators is not unique and there exists infinite number of inequivalent
representations of the basic commutator algebra for the system. 
This shows that 
``physical'' quantities like ground state, particle number etc. will depend on
the specific representation chosen and will not be unique.

(b) Since the system has infinite number of degrees of freedom, quantities
like total energy can diverge. The actual form of the
divergence depends on the representation chosen for the algebra and the
differences between infinite quantities may retain a representation dependent
[finite] value, unless one is careful in regularising such expressions. In
some cases, one may be forced to choose particular set of harmonic oscillators
because of the boundary conditions. Then, the difference between
two infinite quantities could  be physically relevant (and even observable as in the case of, for example, Casimir effect).

(c) The situation becomes worse when the perturbation is switched on. In general,
the perturbation series will not converge and has to be interpreted as an
asymptotic expansion. Further, the individual terms in the perturbation series
will not, in general, be finite creating a far more serious problem.  This is related to the fact that virtual quanta
of {\it arbitrarily} high energy are allowed to exist in the theory [needed for incorporating Lorentz invariance at arbitrarily small length scales] and
still propagate as free fields.

(d) Perturbation theory completely misses all effects which are nonanalytic
in the coupling constant. In QED, for example, perturbation theory cannot
lead to the result that an external electromagnetic field can produce
$e^+-e^-$ pairs \cite {schwing1951} since this effect has nonanalytic dependency on $e$ through a factor 
 $ \exp\left[-\left(\pi m_e^2/|eE|\right)\right]$.
   
 How does one cope up with these difficulties? Issue (a) is handled by choosing
one particular representation for the free field theory by fiat, and working
with it --- and ignoring all other representations. This also dodges the issue (b)  through normal ordering, once a representation for the harmonic oscillators is chosen. Issue (d) is accepted
as a failure of the method and then
ignored. Most of the successful effort was concentrated on handling the
problem of infinities {\it in the individual terms} of the perturbation series,
that is, on issue (c). The paradigm for handling these infinities can be
stated in terms of 
the concept of renomalization which -- though  has nothing to do with divergences, a priori  -- does allow one to cure the  divergences,
 if {\it all} the divergent terms of a perturbation expansion can be
eliminated by redefining the coupling constants in the theory. For an
arbitrary  field theory, we have no assurance that all the divergences can
be so eliminated; in fact, it is quite easy to construct well defined classical field theories for which divergences cannot be eliminated by this process. 

The unexplained
miracle of quantum field theory lies in the fact that several
physically relevant field theories --- describing quantum electrodynamics,
electro-weak interactions and QCD --- belong to this special class of {\it perturbatively renormalisable} theories. Nobody knows why this mathematically non-rigorous,
conceptually ill-defined, formalism of perturbative quantum field theory
works. The miracle becomes even more curious when we notice that
the bag of tricks fails miserably in the case of gravity.

\section{Gravity: Thorn in the flesh}

Until seventies, most of the hardcore particle physicists used 
to ignore general relativity and gravitation and the first concrete attempts in
putting together  principles of quantum theory and gravity were led by 
general relativists (see e.g. \cite{dewitt67}). It was clear, right from the beginning, that
this is going to be a formidable task since the two ``theories of principle''
differed drastically in many aspects:

(a) The Lagrangian describing classical gravity, treated as a function of
$h_{ik}=g_{ik}-\eta_{ik}$, is {\it not} perturbatively renormalizable;
in fact, there does not exist any simple redefinition of the field variables
which will lead to a perturbatively renormalizable theory. So the most
straight forward approach, based on the belief that nature will continue to be kind
to us, is blocked. The miracle fails.

(b) The principle of equivalence implies that any reasonable description of
gravity will have a geometrical structure and that gravitational field
will affect the spacetime intervals in a specific manner, thereby making
the spacetime itself dynamical. For a general gravitational
field, there will be no way of choosing a special class of spacelike hypersurfaces
or a time coordinate. 

(c) Gravity affects the light signals and hence determines the causal 
structure of spacetime. In particular, gravity is capable of generating
regions of spacetime from which no information can reach the outside
world through classical propagation of signals. This feature, which 
may be loosely called `the existence of trapped surfaces' has no 
parallel in any other interaction. When gravity makes certain regions inaccessible, the data regarding quantum fields in these regions can ``get lost''. This requires reformulation of the equations of quantum field theory, possibly by tracing 
over the information which resides in the inaccessible regions --- something
which is not easy to do either mathematically or conceptually. 

(d) Since all matter gravitates, the gravitational field becomes more and more
dominant at larger and larger scales. In the limit, the asymptotic structure of
spacetime is determined by global, smoothed out distribution of matter 
in the cosmological context. Hence, the spacetime will not be
asymptotically flat in the spatial variables at any given time. The behaviour
of the spacetime for $t\to \pm \infty$ will also be highly
non-trivial and could be dominated by very strong gravitational fields. 

(e) All energies gravitate thereby removing the ambiguity in the zero level
for the energy, which exists in non-gravitational interactions. This feature also suggests that there is no such thing as a free, non-interacting
field. Any non trivial classical field configuration will possess certain amount
of energy which will curve the spacetime, thereby coupling the field to itself
indirectly. Gravitational field is not only nonlinear in its own coupling,
but also makes {\it all matter fields} self-interacting.
  
  These features create problems even when one tries to develop a quantum 
field theory in an external gravitational field. Conventional quantum 
field theory works  best when 
a static causal structure, global Lorentz frame, asymptotic in-out states,
bounded Hamiltonians and the language of vacuum state, particle excitations
etc., are supplied.  The gravitational field removes all these features, strongly hinting that we may be working
with an inadequate language. 
Perturbative language  which --- at best --- gives an algorithm to calculate
S-matrix elements, is not  going to be of much use in understanding the quantum structure of gravitational field. Most of the interesting questions --- possibly {\it all} the interesting questions --- in quantum gravity are non perturbative in character; whether a theory is perturbatively renormalizable or not
is totally irrelevant in this context.
The gradual paradigm shift in the particle physics community from perturbative renormalisability (in 70's)  through perturbative
 finiteness of supergravity (in early 80's) to  non perturbative description of superstrings (in late 90's)  represents
a grudging acceptance of the lessons from gravity.

Finally, one may ask --- given these difficulties --- is it really necessary to quantise gravity ?
The answer is "yes" and can be proved in two steps: (i) One can easily prove that if the Casimir energy does not gravitate, it is possible to construct a perpetual motion machine using two Casimir plates and a set of weights and pulleys (see e.g \cite{tprev89}). (ii) If the source for gravity is quantum mechanical (like Casimir energy) but the field is classical then it is possible to violate the uncertainty principle by a suitable set up. Thus at least some minimal amount of quantum structure need to be imposed on gravity. 

\section{Role of Planck length in the microstructure of spacetime}

Having summarised the points of conflict, let me now turn to the points of contact, beginning with the first of the four themes I mentioned in the Introduction. 
 The fact that all matter gravitates stresses the need to abandon
description based on free field theory to handle virtual excitations with arbitrarily high energies.  An excitation with energy $E$ will probe length scales
of the order of $(1/E)$ and when $E\to E_P$, the  nonlinearity due to self gravity cannot be ignored for any field. The same conclusion is applicable even to vacuum fluctuations of any field, including gravity. If we attempt to treat the ground state
of the gravitational field as the flat spacetime, we must conclude that the
spacetime structure at $L \la L_P$ will be dominated by quantum fluctuations of gravity and the smooth macroscopic spacetime can only emerge when the 
fluctuations are averaged over larger length scales.  Hence the description of continuum spacetime in terms of, classical, Einstein's
equation should be thought of as  similar to the description of a solid by elastic constants. While the knowledge
of microscopic quantum theory of atoms and molecules will allow us, in principle, to construct the description in terms of elastic constants, the reverse 
process is unlikely to be unique. What one could hope is to take clues
from well designed thought experiments, thereby identifying some key 
generic features of the microscopic theory.
 
To begin with, one can prove -- using well-chosen thought experiments -- that it is not possible to
measure intervals smaller than $L_P=(G\hbar/c^3)^{1/2}$. 
(This is demonstrated in \cite{tp87}; a clear statement in this direction, based on a toy model, is in \cite{dewitt64} and a more ``modern'' approach to the same result is in \cite{amoti89}.) More formally,
one can prove that the quantum fluctuations in the metric lead to the following limit (see ref.\cite{tp87}, \cite{tp2631}.)

\bee
\qquad\qquad\lim_{ x \to y} <l^2(x,y)>\approx (x-y)^2  + L_P^2 \label{eqn:one}
\eeq 
where $l$ is the geodesic distance between $x^i$ and $y^i$ and the averaging is over all metric fluctuations. This suggests that Planck length should be thought of as the ``zero-point length'' of the spacetime and any correct theory of quantum gravity must incorporate this feature in a suitable form. 

One specific consequence of this result is in the case of a Friedmann model for the universe. It can be shown
that \cite{TP82} the lower bound at Planck length leads to an effective metric of the form

\bee
ds^2=dt^2 - L_P^2(n+{1\over 2})\left[d\chi ^2+ \sin^2\chi (d\Omega^2)\right]
\label{eqn:two}\eeq
leading to the {\it areas} of spherical surfaces being quantised in units of $L_P^2$.

It is {\it not} possible to reconcile the the existence of a zero point length in (\ref{eqn:one}) with a 
Lorentz invariant, local, QFT description. Both string
 theory and loop gravity (the two approaches which have been worked out to fair degree of detail) incorporate this lower bound in different ways. In string theory, nonlocality is built in and hence it is possible to obtain Lorentz invariance at the long-wavelength limit. The situation is less clear in loop gravity but it has the most direct implementation of this principle in, for example, area quantisation. In loop gravity
the area operator is quantised but the eigenvalues scale differently \cite{abhay92} compared to (\ref{eqn:two}). More recently, there has been attempts to construct quantum cosmological models based on loop gravity (see e.g. \cite{mbojo01}) and it appears that
results like (\ref{eqn:two}) might arise as an asymptotic limit.

I have attempted \cite{tp97} to use the interpretation of $L_P$ as a zero-point length to provide a working description of quantum field theory which is free of ultraviolet divergences. The
starting point of this analysis is to modify the path integral for Euclidean Green function $G_F(x,y)$  in such a way that
the path integral amplitude is invariant under the ``duality'' transformation $l\to (L_P^2/l)$ where $l^2=(x-y)^2$. This demands the
replacement
\bee
G(x,y)=\sum e^{-l(x,y)}\to G_{\rm modified} \equiv\sum\exp[-
(l +{L_P^2\over l})]\label{eqn:three}
\eeq
Remarkably enough, it turns out that the path integral sum in (\ref{eqn:three}) can be evaluated rigorously. The final result is quite simple: the modified Green function is related to the original one by the
replacement $(x-y)^2\to [(x-y)^2+L_P^2]$ ! That is,
$ G_{\rm modified}(x,y)=G_{\rm usual}[l^2\to (l^2+L_P^2)]$
In other words, the postulate of  duality (as defined above) implies the existence of a zero-point length. It is known that string theories -- which have zero-point length built in -- do lead to dualities of different kinds. I would like to stress that there is no simple reason to expect, a priori,
 a connection between (\ref{eqn:three}) and zero-point length. 

Once the postulate of path integral duality is accepted, it is possible to obtain several interesting consequences: (i) To begin with, it is clear that gravity acts as a {\it non perturbative} regulator. For example, the modified Feynmann propagator
for a massless scalar field has the structure (see \cite{dewitt64},\cite{tp97}):
\bee
{1\over x^2}\to {1\over (x^2 + L_P^2)}={1\over x^2} -{L_P^2\over x^4} + \cdots 
\eeq
Each term on the right hand side diverges as $x \to 0$ and only the sum remains finite  in the coincidence limit, 
as $x \to 0$.
(ii) One can compute the corrections to the bare cosmological constant $\Lambda_{\rm bare}$  and 
Newtonian  gravitational  constant  $G_{\rm bare}$ when other fields   are integrated out in a path integral. One finds that
\bee
 \Lambda_{\rm ren} = \Lambda_{\rm bare} - {1\over 4\pi \eta^4 G}; \quad 
 G_{\rm ren}^{-1} = G_{\rm bare}^{-1} \left[ 1 + {1\over 12 \pi \eta^2}\right]
\eeq
where $\eta$ is a pure number related to the number of scalar fields integrated out. The result shows that the value
{\it $\Lambda=0$ is unprotected} against large quantum gravitational corrections. (iii) Any form of area quantisation implies that the  density of BH states on the horizon is of the order of $(A/L_P^2)$ with clear implications for the entropy of blackhole. (iv) The zero point length also suggests that there will be exponential suppression of modes shorter than Planck length. This, for example, will allow inflation at quantum gravitational scales
\cite{TP88} since the production of gravitational waves will be suppressed. 

\section{Why did the universe become classical ?}
  
Let me now turn to the second theme, viz. quantum cosmology.
Considering the fact that quantum cosmology was one of the earliest points of contact between QT and GR, it is rather disappointing that it has not produced any concrete results. 
Fundamental questions regarding the origin of the universe remain 
 unanswered in all approaches and even the descriptive language requires semiclassical crutches.
Serious technical questions (eg, exact validity of minisuperspace, canonical vs
path integral approaches, Euclidean vs Lorentzian path integral, topology change .....) are still controversial.

Most of the early work in quantum cosmology was on the question of singularities (and ``creation"
of the universe) and these models (see eg.\cite {misner69}, \cite{hh83}, \cite{TP82}, \cite{TP1718})
 were the precursors of the currently more fashionable
[though hardly better justified] pre-bigbang models. It is fairly easy to construct toy quantum cosmological models  without singularity or horizon. One example I worked out long back \cite{TP1718} has the effective metric: 
\bee
 ds^2 \cong \left( \alpha L_P^2 +\tau^2 \right) \left[ d\tau^2 - dx^2 - dy^2 - dz^2\right]; \quad \alpha = {\cal O} (1)
\eeq
The difficulty with such pre-big bang models is that they are hopelessly diverse
and do not give any more insight than the original assumptions of the model. 

Somewhat more concrete results exist as regards the semiclassical limit of the quantum cosmology, especially since we cannot invoke an ``observer' in this context.
It is possible to show that decoherence provides an answer and leads to the density matrix of 3-geometries becoming effectively diagonal. One can  define a ``distance" in the superspace of 3-geometries $l^2 ( g_{\alpha\beta},\  g'_{\alpha\beta})$ in terms of which one can illustrate \cite {tp89} the suppression of off-diagonal components of the density matrix explicitly:
\bee
\hskip2em\rho_{\rm off-diagonal} \approx \rho_{\rm diagonal} \ \exp\left[ - {l^2 ( g_{\alpha\beta},\  g'_{\alpha\beta})\over L_P^2}\right] 
\eeq

There is another interesting aspect \cite{tp126} to this analysis: Among all systems dominated by gravity, the universe
possess a very peculiar feature. If the conventional cosmological models are 
reasonable, it then follows that {\it our universe proceeded from quantum mechanical behaviour to classical behaviour in the course of dynamical evolution
defined by some intrinsic time variable}.  
 In terms of
Wigner functional $W(g,p)$, this transition can be stated as evolution leading to, 
 
\bee
W(g_{\alpha\beta}, p^{\alpha\beta})$ = $A(g) B(p) \quad \to \quad W(g_{\alpha \beta, } p^{\alpha \beta}) = F \left[p - p_{\rm class}(g)\right]
\eeq 
where $A, B$ are arbitrary functionals and $F$ is a functional sharply peaked on its argument. This transition is {\it not} possible for systems with bounded Hamiltonians arising in
a low-energy effective theory with finite number of fields integrated out. It follows that the quantum cosmological description  of our universe, as a Hamiltonian system, should contain at least one unbounded degree of freedom. In simple quantum cosmological models, one can write
${\cal H} = {\cal H}_{\rm unbound} (a) + {\cal H}_{\rm bound} (a,q); $
where $a$ is the expansion factor and $q$ denotes all other degrees of freedom. It can also be shown that the unbounded mode --- which, in the case of FRW universe, corresponds to the expansion factor $a(t)$ --- will become classical first, as is experienced in the evolution of the universe.
 
One might assume that the  microscopic description of spacetime is in terms of certain [as yet unknown] variables $q_i$ and that the  conventional spacetime metric 
is obtained from these variables in some suitable limit. Such a process will necessarily involve coarse-graining over a class of microscopic descriptors of geometry.
    If one starts with a bounded Hamiltonian for a system with {\it finite}  number of quantum fields and integrate out a   subset of them,
 the resulting Hamiltonian for the low energy theory cannot be unbounded. 
Assuming that the original theory is describable in terms of a bounded Hamiltonian for some suitable variables, it follows that an infinite number of fields have to be involved
in its description and an infinite subset of them have to be integrated out
in order to give the standard low energy gravity. This feature is indeed present in one form or the other in the descriptions of quantum gravity based on
strings or loop variables.   
 
These arguments can be extended further. Starting from an (unknown) quantum gravitational model, one can invoke 
a sequence
of approximations to progressively arrive at quantum field theory (QFT) 
in curved spacetime, QFT in flat spacetime, 
nonrelativistic quantum mechanics and Newtonian mechanics. 
The more exact theory can put restrictions on the range of possibilities
allowed for the approximate theory which are {\it not derivable from  the latter } -- an example being the symmetry restrictions on the wave function for a pair
of electrons in point QM which has its origin in QFT. The choice of vacuum state at low energies could
be such a ``relic'' arising from combining the principles of quantum theory 
and general
relativity \cite{TP00}. The detailed analysis suggests that the wave function of the universe, 
when  describing the large volume limit of the universe, 
dynamically selects a vacuum state for matter fields --- which, in 
turn, defines the concept of particle in the low energy limit.
The result also has the 
potential for providing a concrete quantum mechanical version of 
Mach's principle.

 
\section{Thermodynamics and/of geometry: Can cosmological constant evaporate?}

One of the remarkable features of classical gravity is that it can wrap up regions of spacetime thereby producing surfaces which act as one way membranes. The classic example is that of a Schwarzschild blackhole which has a compact surface
acting as observer independent event horizon. Another example is the deSitter universe which also has an one way
membrane;  but the location of the horizon depends on the observer and hence is coordinate dependent. In fact, the existence of one-way membranes is not necessarily a feature of curved spacetime;
it is possible to introduce coordinate charts even in the flat  Minkowski spacetime, such that regions are separated by horizons. The familiar
 example is  the Rindler coordinate frame which has a non-compact surface acting as a coordinate dependent horizon.

All the three spacetimes mentioned above (Schwarzschild, deSitter, Rindler)
as well as a host of other spacetimes with horizons can be described in a general manner as follows. Consider a $(D+1)$-dimensional {\it flat} Lorentzian
manifold with the line element

\bee
ds^2 = (dZ^0)^2 - (dZ^1)^2 ....(dZ^D)^2 \equiv \eta_{AB} dZ^AdZ^B \label{eqn:nine}
\eeq
Spacetimes of relevance to us can all be thought of $4-$dimensional sub manifolds
of this $(D + 1)-$dimensional manifold, defined in a suitable manner. I will first
introduce two new coordinates $(t,r)$ in place of $(Z^0,Z^1)$ through the definitions

\bee
Z^0 = lf(r)^{1/2} \sinh \ \ gt; \quad Z^1 =  \pm lf(r)^{1/2} \cosh gt \label{eqn:ten}
\eeq
where $(l,g)$ are constants introduced for dimensional reasons and we will usually take $l \propto (1/g)$.  Clearly the pair of points $(Z^0,Z^1)$ and $(-Z^0,-Z^1)$ are mapped to the same
$(t,r)$ making this a 2-to-1 mapping.
(The transformations in (\ref{eqn:two}) covers only the  
two quadrants  with $|Z^1|> |Z^0|$ with the positive sign
for the right quadrant and negative sign for the left but can easily be extended to other quadrants with $\sinh$ and $\cosh$ interchanged).
Note that if one introduces the Euclidean continuation of
the time coordinates with $iZ^0\equiv T;it=\tau$, the transformations in (\ref{eqn:two})  continue to be
valid but with a periodicity of $(2\pi/g)$ in $\tau$.   
With the transformation in (\ref{eqn:two}), the metric in (\ref{eqn:nine}) becomes

\bee
ds^2 = f(r)(lg)^2dt^2 - {l^2 \over 4} \left({ f^{\prime 2} \over f}\right) dr^2 - dL^2_{D-1} \label{eqn:four}
\eeq
 in all the four quadrants. The choice of $D$ and the definition of the four dimensional subspace depends on
the spacetime we are interested in: (a) In the simplest case of Rindler spacetime we can take  $f=(1+2gr), l=g^{-1}$ with all the (D-2) transverse dimensions going  for a ride. In fact, we can treat this case as  just a redefinition of coordinates, involving a mapping from $(D+1)=4$ to $(D+1)=4$.  (b) If we take
$(D+1)=5$ and use --- in addition to the mapping given by (\ref{eqn:two}) --- a transformation
of Cartesian $(Z^2,Z^3,Z^4)$ to the standard spherical polar coordinates: $(Z^2, Z^3, Z^4) \rightarrow (r, \theta, \varphi)$,
and choose  $lg = 1 ; f(r) = \left[1 - (r^2/ l^2) \right]$, we get the deSitter spacetime in static coordinates. (c) To obtain the Schwarzschild spacetime\footnote[{2}] {This was first obtained in \cite{fronsdal59} but the analysis in this reference hides the simplicity and generality of the result.} we  start with a $(D + 1)=$
6-dimensional spacetime  $(Z^0,Z^1,Z^2,Z^3,Z^4,Z^5)$ and consider a mapping to
4-dimensional subspace in which: (i) The  $(Z^0,Z^1)$ are mapped to $(t,r)$ as before; (ii) the 
$(Z^3,Z^4,Z^5)$ are mapped to standard spherical polar coordinates: $(r, \theta, \varphi)$ {\it and} (iii) we take $Z^2$ to be an arbitrary function of $r$: $Z^2=q(r)$. This leads to the metric
\bee
ds^2 = A(r) dt^2 - B(r) dr^2 - r^2 d \Omega^2_{\rm 2-sphere}; \label{eqn:Ttena}
\eeq
with

\bee
A(r) = (lg)^2f; \qquad B(r) = 1 + q^{\prime 2} + {l^2 \over 4} {f^{\prime 2 }\over f} \label{eqn:eleven}
\eeq
This choice will allow us to obtain {\it any} spherically
symmetric, static, 4-dimensional spacetime. For the  Schwarzschild solution  we will take $2lg =1, f=4 \left[ 1 - (l/ r) \right]$; and
\bee
q(r) = \int^r\left[\left( {l \over r}\right)^3 + \left({l  \over r}\right)^2 + {l \over r}\right]^{1/2} dr \label{eqn:thirteen}
\eeq
Though the integral cannot be expressed in terms of elementary functions, it
is obvious that $q(r)$ is well behaved everywhere including at $r=l$.  
 
In the examples of spacetimes with horizons,
  $f(r)$ vanishes at some $r=l$ so that $g_{00} \approx |(r/l -1)|$
near $r=l$; such spacetimes have a horizon at $r=l$. There exists a natural definition of QFT in the original (D + 1)-dimensional space and
we can define a vacuum state for the quantum field on the $Z^0=0$ surface,
which coincides with the $t=0$ surface. It is straightforward to show that this vacuum state
appears as a thermal state with temperature $T=(g/2\pi)$ in the 4-dimensional subspace. The most important conclusion which follows from this analysis \cite{tpwork} is that the existence of the temperature is a purely  kinematic effect arising from the coordinate system we have used --- which should also be obvious from the fact that (\ref{eqn:ten})  implies periodicity in imaginary time coordinate.

The QFT based on such a state will be manifestedly time symmetric and will describe an isolated system in thermal equilibrium in the subregion ${\cal R}$. No time asymmetric phenomena like evaporation, outgoing radiation, irreversible changes etc can take place in this situation. This is
gratifying since one may be hard pressed to interpret an evaporating Minkowski spacetime or even an observer dependent evaporation of a deSitter spacetime; by choosing to work with the quantum state which is time symmetric we can bypass such conceptual issues. 

It is also possible to show that 4-dim QFT gets mapped to 2-dim Conformal Field Theory (CFT)  in all these spacetimes. Consider, for example,  a QFT for a self-interacting scalar field in
a spacetime with the metric of the form 
in equation (\ref{eqn:four}):  

\bee
ds^2 = f(r)(lg)^2dt^2 - {l^2 \over 4} \left({ f^{\prime 2} \over f}\right) dr^2 - g_{\alpha \beta} dx^{\alpha}dx^{\beta}; \quad g_{\alpha \beta} = g_{\alpha \beta} (r, {\bf x}_{\bot}) \label{eqn:dumone }
\eeq
where the line element $g_{\alpha \beta} dx^{\alpha}dx^{\beta}$ denotes the irrelevant transverse part corresponding to the transverse coordinates
${\bf x}_\perp$, {\it as well as} any regular part of the metric
corresponding to $dr^2$. The
field equation for a scalar field in this metric can be reduced to the form 
\bee
{\partial^2 \phi \over \partial t^2} - {\partial^2 \phi \over \partial \xi^2}  = \left( {2g\over f'}\right)^2 {f^2\over Q} \left({\partial \phi\over \partial r}\right) \left({\partial Q\over \partial r}\right) +  (lg)^2 f \left[ (\nabla^2_{\bot}\phi) + {\partial V \over \partial \phi} \right]
\eeq
where $2g\xi=lnf$. The right hand side vanishes as $r\to l$ because $f$ vanishes faster than
all other terms. It follows that near the horizon we are dealing with
a (1+1) dimensional CFT  governed by
\bee
{\partial^2 \phi \over \partial t^2} - {\partial^2 \phi \over \partial \xi^2}
\approx 0 
\eeq
which has an extra symmetry of 
conformal invariance.
 
If we take $f(r)\propto  (r-l)$ near $r=l$ and separate the time dependence by
$\phi = \phi_\omega e^{-i\omega t}$, it is easy to see that -- near
$r=l$ -- the fundamental wave modes are $ 
\phi = e^{-i\omega t \pm i \omega \xi} = e^{-i \omega (t \pm \xi)}  
 = \left( e^{-i \omega z}, e^{-i\omega \bar z } \right)\label{eqn:thirty}
$ where $\tau = it$ and $z \equiv (\xi + i \tau)$ is the standard complex coordinate of the conformal field theory. The boundary condition on the horizon can be expressed most naturally in terms of $z$ and $\bar z$. For example, purely ingoing modes are characterised by $(\partial f/\partial \bar z)=0; (\partial f / \partial z) \not= 0$. Since the system is periodic in $\tau$, the coordinate $z$ is on a cylinder
$(R^1\times S^1$) with $\tau$ being the angular coordinate ($S^1$) and $\xi$ being the $R^1$ coordinate. 
The periodicity in $\tau$ is clearer if we introduce the related complex variable $\rho$ by the definition $ \rho = \exp g (\xi + i\tau) = \exp(gz)$. The coordinate $\rho$ respects the periodicity in $\tau$ and is essentially a mapping from a cylinder to a plane.
It follows that the modes $(e^{-i \omega z}, e^{-i\omega \bar z})$ become $(\rho^{-i\omega/ g}, \bar\rho^{-i\omega/ g})$ in terms of $\rho$.
 
The situation is simpler in the case of a free field with $V=0$. Then
one can show that, near the horizon, $r \simeq l, r^{\prime} \simeq l$ the two point function will have the limiting form
\bee
G\left(t -t^{\prime}; r, r^{\prime} ; {\bf x}_{\bot} , {\bf x}^{\prime}_{\bot} \right)\cong \left\{ \sum\limits_{\lambda} f_{\lambda}({\bf x}_{\bot}) f_{\lambda}({\bf x}^{\prime}_{\bot}) \right\} \left\{ \sum\limits_{\omega} e^{-i \omega[(t-t^{\prime})\pm(\xi - \xi^{\prime})]}\right\} \label{eqn:thirtyseven}
\eeq
where the function $f$ is the eigenfunction of transverse Laplacian with
(set of) eigenvalue(s) $\lambda$;
that is $ \nabla^2 _{D-1} f=-\lambda^2f$. In other words, the two point function factorises into a transverse and radial part
with the radial part being that of a two dimensional massless scalar field. The latter is the same as the Green function of the standard conformal field theory.
 
The role of horizon in producing a CFT can be summarised as follows: In a general $(D+1)$ dimensional theory with $D\geq 2$, we do not have conformal
invariance. If we can kill the transverse dimensions and reduce the theory to a 2-dimensional theory, then we would have automatically enhanced the symmetries to that of a CFT. The metrics in (\ref{eqn:four}) with $g_{00}$ having a simple zero at $r=l$
achieves exactly this. Since such metrics have a horizon, we obtain a connection between CFT and horizons quite generically. All the results of CFT (especially
the behaviour of two-point functions) can now be used to study the field theory near the horizon. 
 
Trapped surfaces
also highlight the role of boundary conditions (called holographic principle in some contexts) in  QFT. The structure of a free
field propagating in an arbitrary spacetime can be completely specified
in terms of, say, the Feynmann Greens function $G_F(x,y)$ which satisfies
a local, hyperbolic, inhomogeneous, partial differential equation. Each solution to this equation provides a particular realization of the theory
so that there exists a mapping between the realizations of the 
quantum field theory and the relevant boundary conditions to this equation which specify a useful solution. When trapped surfaces exists, the differential operator governing the Greens function will be singular on these surfaces 
(in some coordinate chart) and the issue of boundary conditions become far more complex. It is, nevertheless possible --- at least in simple cases with compact trapped surfaces --- to provide
an one-to-one correspondence between the ground states of the theory and the 
boundary conditions for $G_F$ on the compact trapped surface. In fact,
the Greens function connecting events outside the trapped surface can be completely determined in terms of a suitable boundary condition on the trapped surface, indicating that
trapped surfaces acquire a life of their own even in the context of QFT in CST.
In a way, the procedure is reminiscent of renormalisation group approach,  
but now used in real space to integrate out information inside the trapped
surface and  replace it by some suitable boundary condition.

One would next like to know whether one can associate an entropy with these spacetimes in a sensible manner, given that the notion of temperature arises very naturally. Conventionally there are two {\it very different} ways of defining the entropy. In statistical mechanics, the entropy $S(E)$ is related to the 
degrees of freedom [or phase volume] $g(E)$ by $S(E)=\ln g(E)$. Maximisation of the phase volume for systems which can exchange energy
will then lead to the equality of the quantity $T(E)\equiv (\partial S/\partial E)^{-1}$ for the systems. It is conventional to identify this variable as the thermodynamic temperature. In classical thermodynamics, on the other hand, it is the {\it change in} the entropy, which
can be operationally defined via $dS=dE/T(E)$. Integrating this equation will
lead to the function $S(E)$ except for an additive constant which needs to be
determined from additional considerations. 

In the case of time symmetric spacetimes, if one chooses a vacuum state of QFT which is also time symmetric, then there will be no change of entropy $dS$ and the thermodynamic
route is blocked. But the alternative  definition of $S$ --- in terms of certain degrees of freedom --- is possible even in the time symmetric context. Unfortunately, identifying these degrees of freedom is a nontrivial task. More importantly, the QFT described in the last few paragraphs makes absolutely no mathematical distinction between the 
horizons which arise in the Schwarzschild, deSitter and Rindler spacetimes. Any honest
identification of degrees of freedom in conventional QFT will lead to a definition of entropy {\it for all the three cases}. While the blackhole result is acceptable (the horizon being compact and observer independent), the deSitter
spacetime will have an observer dependent entropy (the horizon is compact but
coordinate dependent) and the Rindler frame will have an infinite, observer dependent entropy (since the horizon is non-compact and observer dependent).
While there is voluminous literature on the temperature associated with these spacetimes, there is virtually no clear, published, discussion on the question: {\it Does
the deSitter and Rindler spacetime possess observer dependent entropies, which can be interpreted sensibly?}

There is an alternative point of view which one can take regarding this issue. The Schwarzschild metric, for example, can be thought of as an asymptotic metric arising from the collapse of a body forming a blackhole. While developing the QFT in such a spacetime we need not maintain time reversal invariance for the vacuum state and --- in fact --- it is more natural to choose a state with purely ingoing modes at early times like the Unruh vacuum state. The  study of the QFT in such a spacetime shows that,
at late times, there will exist a thermal, outgoing, radiation of particles
which is totally independent of the details of the collapse. The temperature in this case will be $T(M)=1/8\pi M$, which is the same as the one found in the case of the state of thermal
equilibrium around an ``eternal" blackhole.  In the Schwarzschild spacetime, which is asymptotically flat, it is also possible to associate an energy $E=M$ with the blackhole.
 Though the QFT in CST calculation  was done in a metric with fixed value of energy $E=M$, it seems reasonable to assume that
as the energy flows to infinity at late times, the mass of the black hole
will decrease.  {\it If} we make this assumption --- that the evaporation of black hole will lead to decrease of $M$ --- {\it then} one can integrate the equation $dS=dM/T(M)$ to obtain the
entropy of the blackhole to be $S=4\pi M^2=(1/4)(A/L_P^2)$ where $A=4\pi (2M)^2$
is the area of the event horizon and $L_P=(G\hbar/c^3)^{1/2}$ is the Planck length.  [This integration  can determine the entropy only
upto an additive constant. To fix this constant, one can make the additional assumption that $S$ should vanish when $M=0$. One may think that this assumption is eminently reasonable since the Schwarzschild metric reduces to Lorentzian
metric when $M\to 0$.  But note that in the same limit of $M\to 0$, the temperature of the blackhole diverges !. Treated as a limit of Schwarzschild spacetime, normal flat spacetime has infinite --- rather than zero --- temperature.] 
The procedure outlined above is similar in spirit to the approach of classical
thermodynamics rather than statistical mechanics.

It is rather intriguing that there exist analogues for the collapsing blackhole in the case of deSitter and even Rindler  \cite{tpwork} . The analogue in the case of deSitter
spacetime will be an FRW universe which behaves like a deSitter universe only at late times. (This is probably the actual state of our universe which has become dominated by a cosmological constant in the recent past). Mathematically,
we only need to take $a(t)$ to be a function which has the asymptotic form
$\exp(Ht)$ at late times. Such a spacetime is, in general, time asymmetric and one can choose a vacuum state at early times in such a way that thermal spectrum of particles exist at late times.
Emboldened by the analogy with blackhole spacetime one can also directly construct quantum states (similar to Unruh vacuum of blackhole spacetime) which are time asymmetric, even in the exact deSitter spacetime, with the understanding that the deSitter universe came about at late times through a time asymmetric evolution.  

The analogy also works for Rindler spacetime which is also time symmetric. The standard vacuum state respects this symmetry and we arrive at a situation in thermal equilibrium.  The coordinate system for an observer with {\it time dependent} acceleration will, however, generalise the  standard Rindler spacetime
in a  time dependent manner.  In particular, one can have an observer who was inertial (or at rest) at early times and is uniformly accelerating at late times. In this case an event horizon forms at late times exactly in analogy with a collapsing blackhole. It is now possible to choose quantum states which are analogous to
Unruh vacuum - which will correspond to an inertial vacuum state at early times and will appear as a thermal state at late times. The correspondence with CFT can now be used to compute $\langle T_{ab}\rangle$ in different `vacuum' states to show \cite{tpwork} that radiative flux exists in the quantum states which are time asymmetric analogues of Unruh vacuum state.

But in deSitter or Rindler spacetimes there is {\it no}
natural notion of energy (unlike in blackhole spacetimes which are asymptotically flat).  In fact, it is not clear whether these spacetimes have an ``energy source" analogous to the mass of the blackhole. While the deSitter spacetime is curved and one might consider the cosmological constant to change with evaporation, the Rindler spacetime is flat with (presumably) zero energy. Hence one is {\it forced} to interpret the quantum field theory in these spacetimes   in terms
of a state of thermal equilibrium with constant temperature but no radiation 
(``evaporation"). It is seems correct to conclude that the horizons
{\it always} have temperature but it may not be conceptually straight forward to associate an entropy (or evaporation) with the horizon in all cases.

This might tempt one take the following point of view: In the case of blackholes, one considers the collapse scenario as ``physical" and the natural quantum state is the Unruh vacuum. The notions of evaporation, entropy etc. then follow in a concrete manner. The eternal blackhole (and the Hartle-Hawking vacuum state) is taken to be just mathematical constructs not realised in nature. In the case of Rindler, one may like to think of time symmetric vacuum state as natural treat the situation as one of thermal equilibrium. This forbids using quantum states with outgoing radiation which could make the  Minkowski spacetime to radiate energy -- which seems
unlikely.  

The real trouble  arises for deSitter spacetime which is gaining in popularity. If the spacetime is asymptotically deSitter, should one interpret it as ``evaporating" at late times  with the cosmological constant
 changing with time ?  This will make cosmological constant behave like quintessence models \cite{tpwork}. The energy source for expansion at early times (say, matter or radiation) is irrelevant just as the collapse details are irrelevant in the case of a blackhole. If this is the case, can one sensibly integrate
the $dS=dE/T$ equation and obtain an entropy for deSitter spacetime, even though the spacetime is not asymptotically flat ? And finally, how does one reconcile the fact that the horizon in this case is observer dependent ? These issues are not analysed in adequate detail in the literature and are under study \cite{tpwork}.

The discussion so far was based on the thermodynamic approach to entropy.
One could ask whether it is possible to
 provide
an alternative statistical mechanics interpretation of the entropy [via the equation $S=\ln g$] to complement the  thermodynamical derivation and --- if so --- does it make a distinction between Schwarzschild, deSitter, and Rindler. The simplest case, of course will be that of a black hole. The study of the extensive literature currently available on this topic, with varying points of view, 
 shows that we do {\it not} have a rigorous and
unambiguous interpretation of the entropy of black hole in 
statistical mechanics terms {\it within the context of QFT in CST}. The situation is more unclear in the case of deSitter and Rindler. It is not even clear what are the degrees of freedom one is talking about though the best bet seems to be that they reside  near the surface of the event horizon rather than inside or outside. 

In the context of quantum gravity there have been attempts to relate $S$ to underlying degrees
of freedom of spacetime \cite{rovelli96}, \cite{strom96}, \cite{carlip99}. String theory provides an interpretation of $S$ in
a very special case of an extremal blackhole while the approach based on loop
gravity leads to the proportionality between entropy and the horizon area in a general context though it cannot provide the proportionality constant unambiguously. 
In both these quantum gravitational approaches, certain degrees of freedom are identified and the logarithm of these degrees of freedom leads to an entropy.  These quantum gravitational approaches, unfortunately,  are not of much help in comparing deSitter and Schwarzschild spacetimes. String theory has difficulty in accommodating a positive definite cosmological constant \cite{ewit01} and --- in any case --- the formalism cannot even handle
normal Schwarzschild blackhole  rigorously at present. The loop gravity
approach suffers from the difficulty that --- while it attributes an entropy to
{\it any} horizon --- the derivation is kinematical in the sense that there could be
selection rules in the theory which have a bearing on emission of radiation. Until these are incorporated, it is not possible to proceed from the entropy to temperature. In short, QG models obtain an entropy but not temperature while QFT in CST can lead to temperature but not to entropy in
a straightforward manner.
  
There is another intriguing connection between trapped surfaces and quantum gravity. 
I have given detailed arguments elsewhere  \cite{tp121124} to show that the event horizon of a Schwarzschild blackhole acts as a magnifying glass, allowing us to probe
Planck scale physics. Consider, for example, a physical system described by a low energy Hamiltonian, $H_{\rm low}$. By constructing a blackhole made from the system with this Hamiltonian and requiring that the blackhole should have a density of states that is immune
to the details of the matter of which it is made, one can show
that the Hamiltonian, $H_{\rm true}$  describing the interactions of the system at transplanckian energies must be related to $H_{\rm low}$ by 
$ H_{\rm true}^2 = \alpha E_P^2 \ln [1 + (H^2_{\rm low} / \alpha E^2_p)]$
where $\alpha$ is a numerical factor. Of course, the description at transplanckian energies cannot be in terms of the 
original variables in the rigorous theory. The above formula should be interpreted as giving the mapping between an effective field theory (described by $H_{\rm true}$) and a conventional low energy theory (described by $H_{\rm low}$) such that the blackhole entropy will be reproduced correctly. 

In fact, one can do better and construct a whole class of effective field theories \cite{tp121124} such that the one-particle excitations of these theories possess the same
density of states as a Schwarzschild blackhole. All such effective field theories are non local in character and possess a universal two-point function
at small scales. The nonlocality appears as a smearing of the fields over regions of the order of Planck length thereby confirming ones intuition 
about microscopic structures, trapped surfaces and blackhole entropy.

\section{Cosmological constant -- to be or not to be}

The last theme I want to mention is the  issue of cosmological constant which is the deepest question that  confronts any attempt to combine the principles of general relativity and quantum theory. If the current observational evidence -- suggesting the existing of a small but 
nonzero cosmological constant -- does not go away, theoreticians have a serious problem in their hands. Let me briefly review the difficulties and possibilities.

To begin with, cosmological constant is {\it not} a problem in classical general relativity.
Classical physics has constants which stay as constants and one is allowed to give any value to them. If we write the Einstein's equations as $G_{ik}+\Lambda g_{ik}=8\pi G_N T_{ik}$ we are free to choose any value we like for the two constants $(G_N,\Lambda)$. Further, one cannot construct any quantity with dimension of $\Lambda$ from $G_N$ and $c$
alone so there is no question of fine tuning. 

The situation changes in three respects
when one brings in quantum theory. (i) With $(G_N,c,\hbar)$ one can construct a dimensionless combination $\Lambda L_P^2$ and one may be justifiably curious why this quantity is
somewhat small --- being less than $10^{-120}$. (ii) The coupling constants in quantum theory ``run". In any
sensible model, with a UV cutoff around Planck scale, the value of $\Lambda L_P^2$ 
will run to a number of order unity; that is, a tiny value is unnatural in the technical sense of the term. (iii) Any finite vacuum energy density $V_0$, including the constants added to potential energy terms of scalar fields, say, will contribute a term $(-V_0 g_{ik})$
at the right hand side of Einstein's equations. This is mathematically indistinguishable from the cosmological constant. It is not clear why $ V_{0,net} L_P^2$ is less than $10^{-120}$. The situation is aggravated by the fact that we do not know of any symmetry
which requires $ V_{0,net} L_P^2$ to be zero.

Ever since recent cosmological observations suggested the existence of a nonzero cosmological constant, there has been a flurry of theoretical activity to `` explain "
it, none of which even gets to the first base. One class of models invokes some version of anthropic principle; but since anthropic principle never predicted anything, I do not
consider it part of scientific methodology. The second class of models use a  scalar field with an ``appropriate" potential $V(\phi)$ to ``explain" the observations. These
models are all trivial and have no predictive power because it is always possible to choose a $V(\phi)$ to account for any sensible dynamical evolution of the universe. Since
the triviality of these models (which are variously called ``quintessence", ``dark energy" ....) does not seem to have been adequately emphasised in literature, let me briefly comment on this issue \cite{blois}.

Consider any model for the universe with a {\it given}
$a(t)$ and some known forms of energy density $\rho_{\rm known} (t)$ (made of radiation,
matter etc) both of which are observationally determined. It can happen that this
pair does not satisfy the Friedmann equation for an $\Omega=1$ model. To be specific,
let us assume $\rho_{\rm known}<\rho_c$ which is substantially the situation in cosmology today. If we now want to make a consistent model of cosmology with $\Omega=1$, say,
we can invoke a scalar field with the potential $V(\phi)$. It is trivial to choose
$V(\phi)$ such that we can account for {\it any} sensible pair [$a(t), \rho_{\rm known} (t)] $ along the following lines: Using the given $a(t)$, we define two quantities
$ H(t)=(\dot a/a)$ and  $Q(t)\equiv 8\pi G \rho_{\rm known}(t) /3H^2(t)$. The required $V(\phi)$ is given parametrically by the equations:
 
\bee
V(t) = (1/16\pi G) H (1-Q)\left[6H + (2\dot H/H) - (\dot Q/1-Q)\right]
\eeq
 
\bee \phi (t) = \int dt \left[ H(1-Q)/8\pi G\right]^{1/2} \left[\dot Q/(1-Q )- (2\dot H/H)\right]^{1/2}
\eeq
All the potentials invoked in the literature are special cases of this formula \cite{blois}. This result shows that {\it irrespective of what the future observations reveal about $a(t)$
and  $\rho_{\rm known}(t)$} one can always find a scalar field which will ``explain"
the observations. Hence this approach has no predictive power. What is worse, most of the $V(\phi)$
suggested in the literature have no sound particle physics basis and --- in fact ---
the quantum field theory for these potentials are very badly behaved on nonexistent.

It is  worth realising that the existence of a
 non zero cosmological constant will be a statement of fundamental significance and constitutes a conceptual contribution of cosmology to quantum gravity. The tendency of some cosmologists to treat $\Omega_\Lambda$
as one among a set of, say, 17 parameters [like $\Omega_{rad}, \Omega_B, n, ....$]
which need to be fixed by observations, completely misses the point. Cosmological constant is special and its importance transcedends cosmology.

At present we do not have a fundamental understanding of cosmological constant from any approaches to quantum gravity. There are no nontrivial string theoretical models incorporating $\rho_V>0$; loop gravity can incorporate it but does not throw any light
on its value. It should  be stressed that the nonzero value for
$\rho_V\neq 0$ does {\it not} imply deSitter (or even asymptotically deSitter)
spacetime. Hence the formalism should be capable of handling $\rho_V$ without deSitter geometry.
   
To give an example of a more fundamental way of thinking about cosmological constant, let me describe an idea in which cosmological constant is connected with the microstructure of spacetime. In this model we start with $\Lambda=0$ but generate a small value for this
parameter from two key ingredients:
 (i) discrete spacetime structure at Planck length and (ii) quantum gravitational
uncertainty principle. To do this, we first note that
cosmological constant can be thought of as a lagrange multiplier for proper volume
of spacetime in the action functional for gravity:
\bee
A_{grav}={1\over 2L_P^2}\int d^4x R\sqrt{-g}-{\Lambda\over L_P^2}\int d^4x \sqrt{-g}; \label{eqn:xyz}
\eeq
In any quantum cosmological models which leads to large volumes for the universe, phase of
the wave function will pick up a factor of the form
$\Psi\propto \exp(-i(\Lambda/L_P^2){\cal V})$, where ${\cal V}$ is the four volume, from the second term in (\ref{eqn:xyz}).
Treating $(\Lambda/L_P^2,{\cal V})$ as conjugate variables $(q,p)$, we can invoke the standard uncertainty principle to predict
 $\Delta\Lambda\approx
L_P^2/\Delta{\cal V}$. Now we make the crucial assumption regarding the microscopic structure of the spacetime: Assume that there is a zero point length of the order of $L_P$
so that the volume of the universe is made of several cells, each of volume $L_P^4$. Then  ${\cal V}=NL_P^4$, implying a Poisson fluctuation $\Delta{\cal V}\approx
\sqrt{{\cal V}}L_P^2.$  and leading to 
\bee
\Delta\Lambda={L_P^2\over \Delta{\cal V}}={1\over\sqrt{{\cal V}}}\approx H_0^2
\eeq
which is exactly what cosmological observations imply!
Planck length cutoff (UV limit) and volume of the universe (IR limit) combine to give the correct $\Delta\Lambda$. Of course, this makes $\Lambda$ a stochastic variable and
one needs to solve Friedmann equations using a stochastic source \cite{tpwork}.




\end{document}